\documentclass[aps,aps,groupedaddress]{revtex4-1}

\newcommand\coord[3]{\begin{pmatrix}
 #1 \\
 #2 \\
 #3
 \end{pmatrix}}
 
\usepackage{xcolor}
 
\newcommand{\bu}{{\bf u}} 
 
\newcommand{\bF}{{\bf f}}
\newcommand{\bk}{{\bf k}}

\newcommand{\bx}{{\bf x}}

\newcommand{\beq}{\begin{equation}}
\newcommand{\eeq}{\end{equation}}

\usepackage{amsmath}
\usepackage{graphicx}

\begin{document}


\title{Role of the forcing dimensionality in thin-layer turbulent energy cascades}


\author{Basile Poujol}
\email[]{basile.poujol@ens.fr}
\affiliation{D\'{e}partement de G\'{e}osciences, \'{E}cole Normale Sup\'{e}rieure, PSL Research University, Paris, France}

\author{Adrian van Kan}
\email[]{adrian.van.kan@phys.ens.fr}
\affiliation{Laboratoire de Physique de l’Ecole normale supérieure, ENS, Université PSL, CNRS, Sorbonne Université, Université de Paris, F-75005 Paris, France}

\author{Alexandros Alexakis}
\email[]{alexakis@phys.ens.fr}
\affiliation{Laboratoire de Physique de l’Ecole normale supérieure, ENS, Université PSL, CNRS, Sorbonne Université, Université de Paris, F-75005 Paris, France}


\date{\today}

\begin{abstract}
We investigate the transition from forward to inverse energy cascade in 
turbulent flows in thin layers, varying the functional form
of the forcing and the thickness of the layer.
We show that, as the forcing function becomes more three-dimensional, 
the inverse cascade is suppressed and the critical height $h_c$,
where the transition occurs, is decreased.
We study 
the dependence of this critical height on 
a parameter $r$ which measures the dimensionality of the forcing 
and thus construct a phase space diagram in the parameter space
$r-h$. 
We discuss the effect of Reynolds number and domain size.

\end{abstract}


\maketitle




\section{Introduction}
\label{sec:intro}
Turbulence is a process that takes place in many natural and industrial flows. 
A key property of three-dimensional turbulent flows is their ability to generate smaller and smaller scales, transferring energy across scales by a cascade mechanism, so that energy is efficiently dissipated  by viscosity \cite{Frisch}. 
However, for many geophysical flows it has been shown that energy can also cascade to large scales and yield what is called an inverse energy cascade \cite{Byrne2013height, king2015upscale, Izakov2013Venus, young2017forward}. This typically occurs when an external mechanism (e.g. rotation) renders the flow quasi-two-dimensional.  
In two dimensions, the conservation of a second quadratic sign-definite invariant, the enstrophy, has a direct influence on the energy cascade, which is directed towards the large scales \cite{Boffetta2012two}. In quasi-two-dimensional turbulence the cascade process is split and some
energy cascades to large scales, following two-dimensional (2-D) dynamics, while some energy cascades to small scales, following  three-dimensional (3-D) dynamics. Such a situation is referred to as a split or bidirectional cascade in the literature. 
The presence of an inverse energy cascade means that the small scales can have a significant influence on the fluid behavior at large scales. This is important for fluid dynamic models where the smallest scales (including the dissipation scale) are not resolved. If they have any influence on the large scales, it must be parameterized. In geophysical fluid modelling, these parameterizations of turbulence are still based on very coarse physical and empirical laws, mainly because of our poor understanding of turbulence dynamics. It is therefore crucial to understand these energy cascade processes in order to improve fluid modeling.

Several parameters can modify the intensity of the energy cascade and its direction, including 
rotation, stratification, or the presence of a magnetic field. In particular, it was found theoretically, experimentally and numerically that the geometry of the domain could have an influence on the energy cascade in the flow and its direction \cite{smith1996crossover, celani2010, xia2011, gallet2015exact_a, gallet2015exact_b, benavides2017, musacchio2017split, vankan2019, musacchio2019condensate} (see \cite{alexakis2018} for a review). This is especially applicable in layers of finite thickness, where the thickness is a control parameter that can alter the geometry of the domain from 3-D to 2-D by compacting one dimension. Moreover, in \cite{benavides2017} it was found that the transition from forward cascade (3-D behavior) to inverse cascade  (2-D behavior) can be critical, meaning that there is a critical height below which the inverse cascade appears.



Whereas the influence of the fluid geometry on the direction of the energy cascade is well documented, the role of the forcing has not been well studied.  One expects that a 3-D forcing (a forcing that depends on all three directions) will be less efficient at generating an inverse cascade than a 2-D forcing (a forcing that depends on two directions). However, most of the existing numerical work used a two-dimensional, three-component (so-called 2D3C) forcing.  In \cite{alexakis2018b}, where large-scale instabilities in a thin-layer flow were studied, it was found that the onset of large-scale instabilities depends strongly on the dimensionality of the background flow. This background flow is generally strongly dependent on the forcing and therefore this suggests that the dimensionality of the forcing might have a strong influence on the onset of a direct or an inverse cascade.
This is of particular interest since most fluid flows (industrial or natural) occur under a complex forcing, and it is possible that the existence of an inverse cascade strongly depends on the dimensionality of the forcing at the injection scale. Therefore, this study aims at understanding the role of the forcing dimensionality in determining the characteristics of the energy cascade in layers of finite thickness.

The rest of this paper is structured as follows: the physical setup and the methodology of the study, based on numerical simulations, are described in section \ref{sec:method}, and then the results are presented in section 
\ref{sec:results}. The conclusions of this work are given in section \ref{sec:conclusions}.

\section{Method}                                   
\label{sec:method}                                 

\subsection{General setting and forcing function}

We consider the hyperviscous Navier-Stokes equations in a periodic box of dimension $2\pi L$ in the $x$ and $y$ (hereafter called horizontal) directions and of vertical height $2\pi H$ in the direction $z$ 
(the vertical direction). They are given by
\begin{equation}
\frac{\partial \bu}{\partial t} + ( \bu \cdot {\bf\nabla} ) \bu = - {\bf \nabla} p - \nu_4 \Delta^4 \bu - \mu_2 \Delta^{-2} \bu + \bF, \qquad \nabla \cdot \bu =0, \label{eq:NS}
\end{equation}
where $\bu$ is the velocity and $p$ is the pressure.  A hyperviscosity $\nu_4$ of order 4 and a hypoviscosity $\mu_2$ of order 2 were used. Their value was chosen so that the large-scale and the small-scale dissipation are well-resolved, while increasing the range of scales that follow inviscid dynamics (inertial scales). This is a necessary choice since it is very difficult, even with today's super-computing power, to have a turbulent behavior both for the forward and for the inverse cascade.  The forcing $\bF$ 
was chosen to be time-independent and given by
\begin{widetext}
\begin{equation}
\bF = F_0 \cos \left( \frac{r \pi}{2} \right) \coord{+\cos(k_fy)}{-\cos(k_fx)}{0} + F_0 \sqrt{\frac{2}{k_f^2+q^2}} \sin \left( \frac{r \pi}{2} \right) \coord{q\sin(qz)\cos(k_fx)}{q\sin(qz)\cos(k_fy)}{-k_f\cos(qz)(\sin(k_fx)+\sin(k_fy))}.
\end{equation}
\end{widetext}
Here $k_f$ is the forcing wavenumber in the horizontal direction and $q$ is chosen as $q=1/H$, so that it is the smallest wavenumber in the vertical direction. It satisfies $\| \bF \| = F_0$ and $\nabla \cdot \bF = 0$ for any value of $r$ and $q$. For $r=0$ the forcing is 2-D since it does not depend on the vertical dimension $z$ and is equal to its vertical average value $\bF=\overline{\bF}$ (where the over-line stands for vertical average). It takes the form of 2-D vortices of positive and negative vorticity arranged in a checkerboard pattern. 
For $r=1$ the forcing has zero 2-D projection $\overline{\bF}=0$ and we will refer to it as purely 3-D. In this case it takes the form of square convection cells.
The 2-D ($r=0$) and the 3-D ($r=1$) terms are orthogonal, and their relative norms are $F_0 \cos \left( \frac{r \pi}{2} \right)$ and $F_0 \sin \left( \frac{r \pi}{2} \right)$.  Therefore as $r$ is varied from 0 to 1 the the forcing changes from 2-D to purely 3-D. The goal of this study is to estimate the influence of the parameter $r$ on the critical values that separate the direct energy cascade regime from the split (inverse and direct) energy cascade regime.

For this, an ensemble of simulations were run in which the parameter $r$ and the remaining non-dimensional numbers were varied.
The equations are solved using the Geophysical High-Order Suite for Turbulence (GHOST) code, \cite{mininni2011}. 
It solves the three-dimensional Navier-Stokes equations in a periodic box using a pseudo spectral method with a 2/3 aliasing.
For each simulation, the flow is randomly initialized and forced. The simulation is carried on until the energy spectrum stabilizes in a statistically steady state. At this point, the energy injection rate, and the hypoviscous and hyperviscous dissipation rates begin to oscillate around an equilibrium value.  Each simulation is run until a steady state is reached and continued for sufficiently long time, so that we have a good estimate of the time average value of all quantities of interest.

\subsection{Control Parameters and observables}         
\label{CPO}                                             

At steady state, where initial conditions are forgotten, the system is controlled by five independent non-dimensional parameters. The first nondimensional control parameter is given by the measure of the forcing dimensionality $r$ where $r=0$ corresponds to 2-D forcing while $r=1$ corresponds to 3-D forcing, as described in the previous section. The second control parameter is the normalized layer thickness 
\beq 
h=k_fH. 
\eeq 
Another geometrical parameter comes from the normalized layer width $\Lambda=k_f L$. 
Finally, we have a hyperviscous and a hypoviscous Reynolds number that can be defined 
as 
\[
Re_\nu \equiv
\frac{\epsilon_\mathrm{inj}^{1/3}}{\nu k_f^{22/3}}
=(k_f l_\nu)^{-22/3}, \quad 
Re_\mu \equiv \frac{\epsilon_\mathrm{inj}^{1/3}k_f^{14/3}}{\mu}
= (k_f l_\mu)^{14/3},
\]
where $\epsilon_\mathrm{inj}$ is the energy injection rate and $l_\nu$ and $l_\mu$ are the typical dissipation length scales associated with hyperviscosity and hypoviscosity. Their expressions are deduced from scaling arguments in equation (\ref{eq:NS}) and are given by, respectively,
\[
l_\nu =  \epsilon_\mathrm{inj}^{-1/22} \nu_4^{3/22},\quad
l_\mu = \epsilon_\mathrm{inj}^{1/14} \mu_2^{-3/14} .
\]


%
The simulations are labeled as L$x$Re$y$ where $x$ indicates $\Lambda$
and $y$ indicates 
the hyperviscous Reynolds number $Re_\nu$. 
The values of the parameters for each simulation can be found in Table 1. 
The vertical resolution $\Delta z$ was taken equal to the horizontal resolution $\Delta x$, except for very thin layers
where $\Delta x$ was too coarse to solve properly the dynamics in the $z$-direction and a smaller value of $\Delta z$ is used: 
$\Delta z = \mathrm{min} \{ \Delta x , \Delta z_\mathrm{max} \}$ where $\Delta z_\mathrm{max}=\pi H/4$.
For each set of simulations, different values of the normalized thickness $h$ and of the forcing parameter $r$ were used.

\begin{table}[h]
\begin{ruledtabular}
\begin{tabular}{ccccc||cccc }
     & $\Lambda$ &    $Re_\mu$    & $k_f l_\mu$  &  &   & $Re_\nu$          &$2\pi/(k_f \Delta x)$& $1/(k_f l_\nu)$  \\
\colrule
L1   &     8     &$1.5 \cdot 10^4$&    7.9  &     & Re1 &$4.1 \cdot 10^6  $ &   32                &  
6.6 \\
L2   &     16    &$4.9 \cdot 10^5$&   16.6   &    & Re2 &$1.9 \cdot 10^8  $ &   64                &  
13.7 \\
L3   &     32    &$9.8 \cdot 10^6$&   31.5    &   & Re3 &$3.6 \cdot 10^{10}$&   128               &  
27.8\\
\end{tabular}
\caption{Parameter values used for the different simulation cases. Simulation L1Re3 stands for a simulation with parameters from the row L1 of the left table, and the row Re3 of the table on the right. Note that the number of grid points in the $x$-direction is given by $2\pi\Lambda /(k_f \Delta x)$ and thus can be obtained by the product of $\Lambda$ with $2\pi/(k_f \Delta x)$ 
that for the example of L1Re3 gives $N_x= 8 \times 128 = 1024$. 
Note also that $k_fl_\mu \lesssim \Lambda$ that implies that the hypodissipation is sufficient
to prevent the formation of a condensate and that $1/k_fl_\nu \lesssim 2\pi/(3 k_f \Delta x) $
that implies the the dissipation wavenumber $k_\nu=1/l_\nu$ is smaller than the maximum wavenumber 
$k_{max} =2\pi/(3 \Delta x)$ imposed by the numerical grid and the 2/3 de-aliasing rule. }
\end{ruledtabular}
\end{table}

The series of simulations we performed
aims at studying the behavior of the fluid in two different limits :
\begin{itemize}
\item The large-box limit $\Lambda\to \infty, \, Re_\mu \to  \infty$. 
When increasing $\Lambda$, we consistently modify the hypoviscosity $\mu$ so that the ratio between the typical hypodissipation length scale $l_\mu = \epsilon_\mathrm{inj}^{1/14} \mu^{-3/14}$ and the box size remains approximately unchanged. Therefore, the low wavenumber inertial range becomes wider as the box size is increased, and it ensures a good scale separation between the energy injection and dissipation scales without the formation of a condensate. In this limit, the energy dissipation rate at large scales becomes independent of $\Lambda$ and $\mu$ and equal to the inverse energy flux rate.
\item The high Reynolds number limit $Re_\nu \to \infty, \, \Delta x \to  0$. 
When increasing the simulation resolution by a factor of $\beta$, we also decrease the hyperviscosity $\nu_4$ such that the energy dissipation (Kolmogorov) scale $l_\nu =  \epsilon_\mathrm{inj}^{-1/22} \nu^{3/22}$ is decreased by a factor $\beta$ as well. Similarly to the previous case, the high-wavenumber inertial range becomes wider as viscosity is decreased, the effects of finite viscosity on the forcing scales are diminished and the forward energy flux becomes equal to the dissipation rate due to hyperviscosity.
\end{itemize}

These limiting procedures allow us to use the energy injection and dissipation rates as a measure of the forward and inverse cascade amplitudes. The energy injection rate $\epsilon_\mathrm{inj}$, the viscous dissipation rate $\epsilon_\nu$ and the hypoviscous dissipation rate $\epsilon_\mu$ are given by
\begin{equation}
  \epsilon_\mathrm{inj} = \langle \bF \cdot \bu \rangle
,\qquad
  \epsilon_\nu = \nu_4 \langle |\Delta^2 \bu|^2 \rangle
,\qquad
  \epsilon_\mu = \mu_2 \langle  |\Delta^{-1} \bu|^2 \rangle,
\end{equation}
where the brackets stand for the average over the simulation domain and over time in the steady state regime.
The dissipation rate $\epsilon_\mu$ is a measure of the rate at which energy arrives at large scales while $\epsilon_\nu$
gives the rate at which energy arrives at small scales. They satisfy $\epsilon_\mathrm{inj} = \epsilon_\nu + \epsilon_\mu $, 
so it is convenient to write the relative rate that energy arrives at large scales as
\begin{equation}
\alpha \equiv \frac{\epsilon_\mu}{\epsilon_{\mathrm{inj}}}.
\end{equation}
The variable $\alpha$ takes values between $\alpha=0$ to $\alpha=1$ with $\alpha=0$ implying no inverse cascade,
while $\alpha=1$ implies no forward cascade.
 
In addition to the energy dissipation rates, we also define the cylindrically averaged energy spectra as a measure of the distribution of energy among scales as
\begin{equation}
  E(k) = \frac{1}{2\delta k}\sum_{k-\delta k < | \bk_\perp | \le k} | \hat{\bu} (\bk) |^2,
\end{equation}
where $\hat{\bu} (\bk)$ stands for the complex Fourier transform coefficients of the velocity field,  $\bk_\perp=(k_x,k_y,0)$
it the wavenumber projected on the horizontal plane and
$\delta k$ is a small increment for the wavenumber norm here taken to be $\delta k= 1/L$ the smallest non-zero horizontal wavenumber. 
Finally, we also define the spectral energy flux across cylinders as:
\begin{equation}
    \Pi(k) = \langle \bu_k^{<} \cdot ( \bu \cdot \nabla) \bu \rangle
,\qquad \mathrm{where} \qquad
    \bu_k^{<} = \sum_{| \bk_\perp | \le k} \hat{\bu} (\bk) e^{i \bk \cdot \bx}
\end{equation}
which is the flux of energy cascading (forward if positive, inversely if negative) across a cylinder in Fourier space aligned with the $z$ axis and of radius $k$.

\section{Results}                                   
\label{sec:results}                                 

\subsection{Basic flow features}           

\begin{figure}
\begin{center}
\includegraphics[width=17cm]{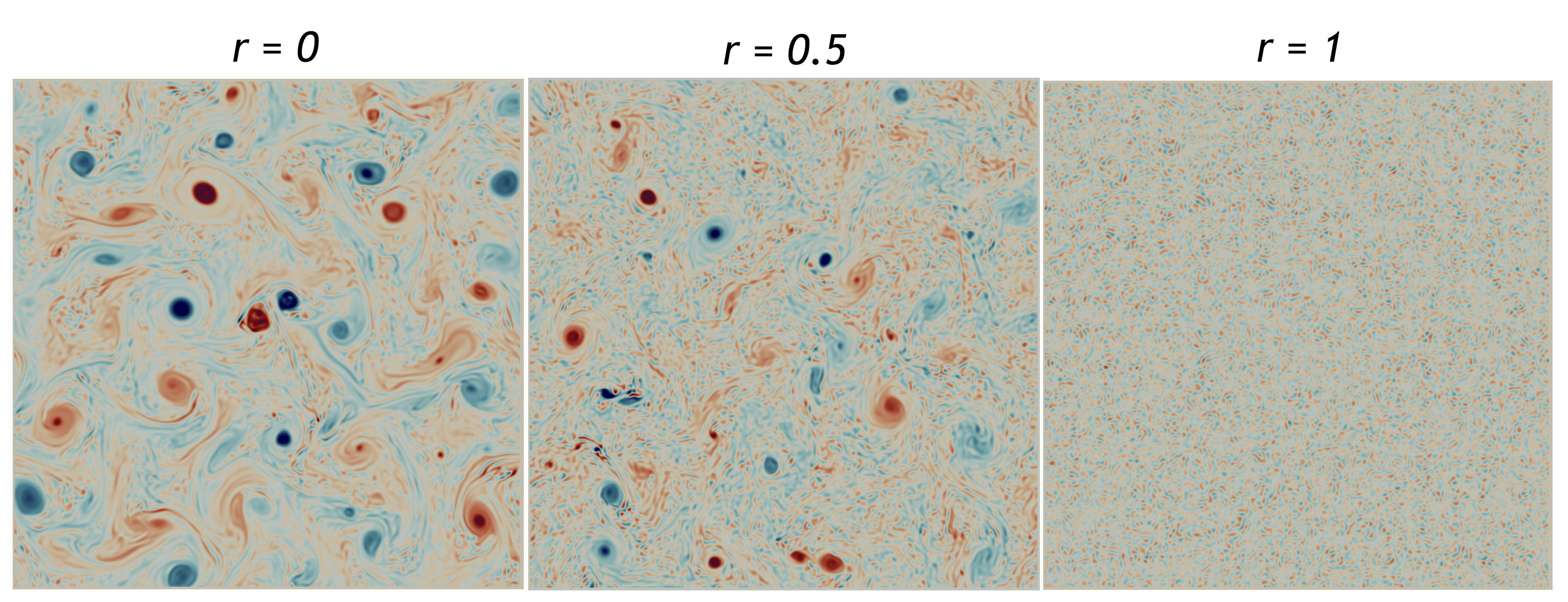}
\end{center}
\caption{ Flow visualization in terms of vorticity for three different values of $r$. 
The color scale is the same for the three subplots with red for positive vorticity and blue negative vorticity. The simulations were done with $h = 1/8$, for the set of simulations L1Re2. \label{fig:vort}}
\end{figure}

We begin by describing the flow in different states of the system. Figure \ref{fig:vort} shows snapshots of vertical vorticity in a horizontal slice of the domain at late times. The results are from the series of runs L1Re2 for different values of $r$ and for $h=1/8$. 
For the purely 2-D forcing $r=0$ (left panel), 2-D vortices are present whose size is close to the forcing scale. These vortices cluster and self-organize as in 2-D turbulence, moving energy to larger scales. The presence of hypoviscosity prevents vortex coalescence and the formation of a condensate vortex of size similar to the box size. 
For  purely 3-D forcing $r=1$ (right panel), such 2D structures are not present. The vorticity is concentrated at scales much smaller than the forcing scale, which is a clear sign that energy is cascading forward towards the small scales, where it is dissipated by hyperviscosity.
Finally, for the intermediate case $r=0.5$ (center panel), we see the superposition of the two states : there are smooth 2-D vortices as in the left panel that coexist with small-scale vorticity structures in between them. This suggests that the energy cascade is bidirectional in this case, as it shares the features of both 2-D and 3-D turbulence. 
It is worth noting that the 2-D features are 
met in some regions, while other regions of space display 3-D features. Thus, it appears that the two processes of forward and inverse energy cascade coexist in the flow, but in different regions of the domain.     

\begin{figure*}
\begin{center}
\includegraphics[width=17cm]{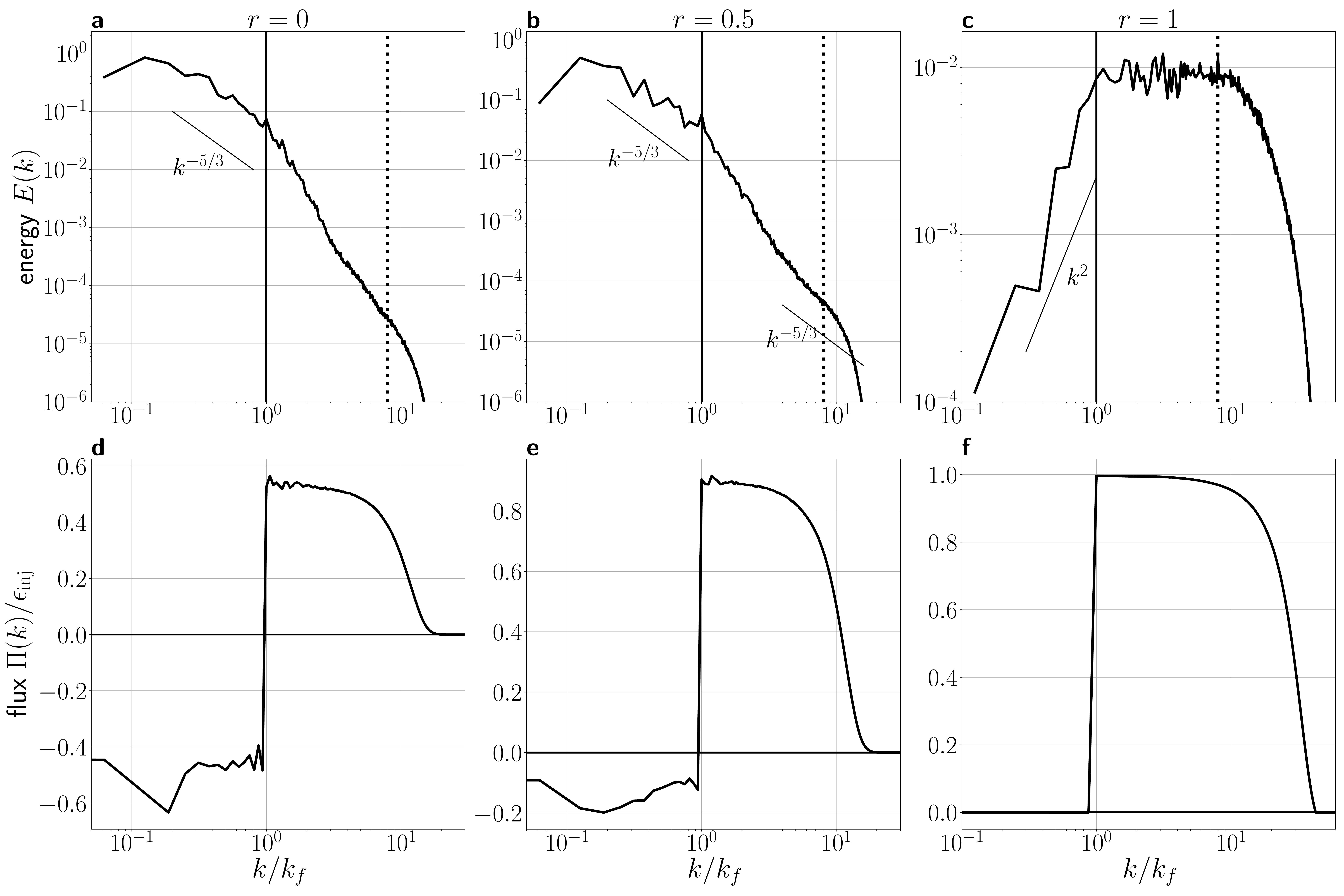}
\end{center}
\caption{(a-c) : Instantaneous energy spectra for different values of $r$, and for $h = 1/8$, after quasi-equilibrium has been reached. Thin straight lines show scalings of $k^{-5/3}$ or $k^2$ as indicated in the annotations. The set of simulation is L2Re2 for the left-hand and center panels, and L1Re3 for the right-hand panel. The dotted vertical line indicates $k = 1/H$. (d-f) : Spectral energy fluxes for the same simulations, averaged over the quasi-equilibrium state. Fluxes are normalized by the energy injection rate $\epsilon_\mathrm{inj}$ 
\label{spectra1}}
\end{figure*}

A more quantitative description of the three different cases can be given by looking at the energy spectra and the energy fluxes of the flows. The top panels of figure \ref{spectra1} display the energy spectra of flows with different values of $r$ and the same value of $h=1/8$ from the series L2Re2 for the left and center panels and from L1Re3 for the right panels. The solid vertical line marks the forcing wavenumber while the dotted vertical line marks the wavenumber $k_h=1/H$ after which 3-D turbulence is expected to be recovered. The bottom panels of the same figures show the corresponding energy fluxes for the same flows as the ones used for the spectra above.

The case $r=0$, shown in the left top panel, is a case that displays an inverse cascade with energy concentrated predominantly in the small wavenumbers. The spectral slope observed at small wavenumbers is consistent with a $k^{-5/3}$ spectrum, although there is not enough range to make a more precise statement. Moreover, the energy flux shown in the bottom left panel shows a strong inverse energy cascade as well. At scales smaller than the forcing scale there is still a forward cascade, but the energy spectrum drops much more steeply.

For $r=1$, shown in the right panels, there is no inverse cascade and 
no accumulation of energy at wavenumbers smaller than the forcing wavenumber. At these scales, we observe a spectrum close to that at absolute equilibrium $E(k) \propto k^2$ \cite{dallas2015statistical, alexakis2019thermal}. The energy spectrum thus peaks at the forcing wavenumber, while at smaller wavenumbers the spectrum is almost flat between $k_f$ and $k_h$. 
At larger wavenumbers  $k \gg k_h$, where the flow behaves as three-dimensional, the spectrum should be proportional to $k^{-5/3}$, but due to grid size limitations the required viscosity is too strong and no Kolmogorov spectrum is observed. However, the energy fluxes are strictly positive in this case, which implies that all the injected energy is cascading forward.

Finally, in the middle panels of the same figure, the spectrum and the flux from a flow with $r=0.5$ are shown. For this case both forward and inverse cascade exist with the energy fluxes, showing the superposition of a weak inverse cascade and a strong forward cascade. The energy peaks at large scales due to the inverse cascade. At wavenumbers larger than $k_f$, there is a steep drop which flattens out as the wavenumber $k_h$ is approached, likely approaching a $k^{-5/3}$ energy spectrum. However, we do not have a sufficiently extended range to ascertain this. 

It is thus clear that the parameter $r$ can change the cascade phase of the systems from forward to bidirectional 
as it is varied, just like the parameter $h$ \cite{benavides2017,celani2010,musacchio2017split}. Thus, in general, the cascade phase of the system depends on both parameters $h,r$, and in order to fully describe its state, a phase space diagram in the ($r,h$) phase space needs to be constructed. This is what we attempt to do in the following subsection.

\subsection{Phase-space diagram}

\begin{figure*}
\begin{center}
\includegraphics[width=17cm]{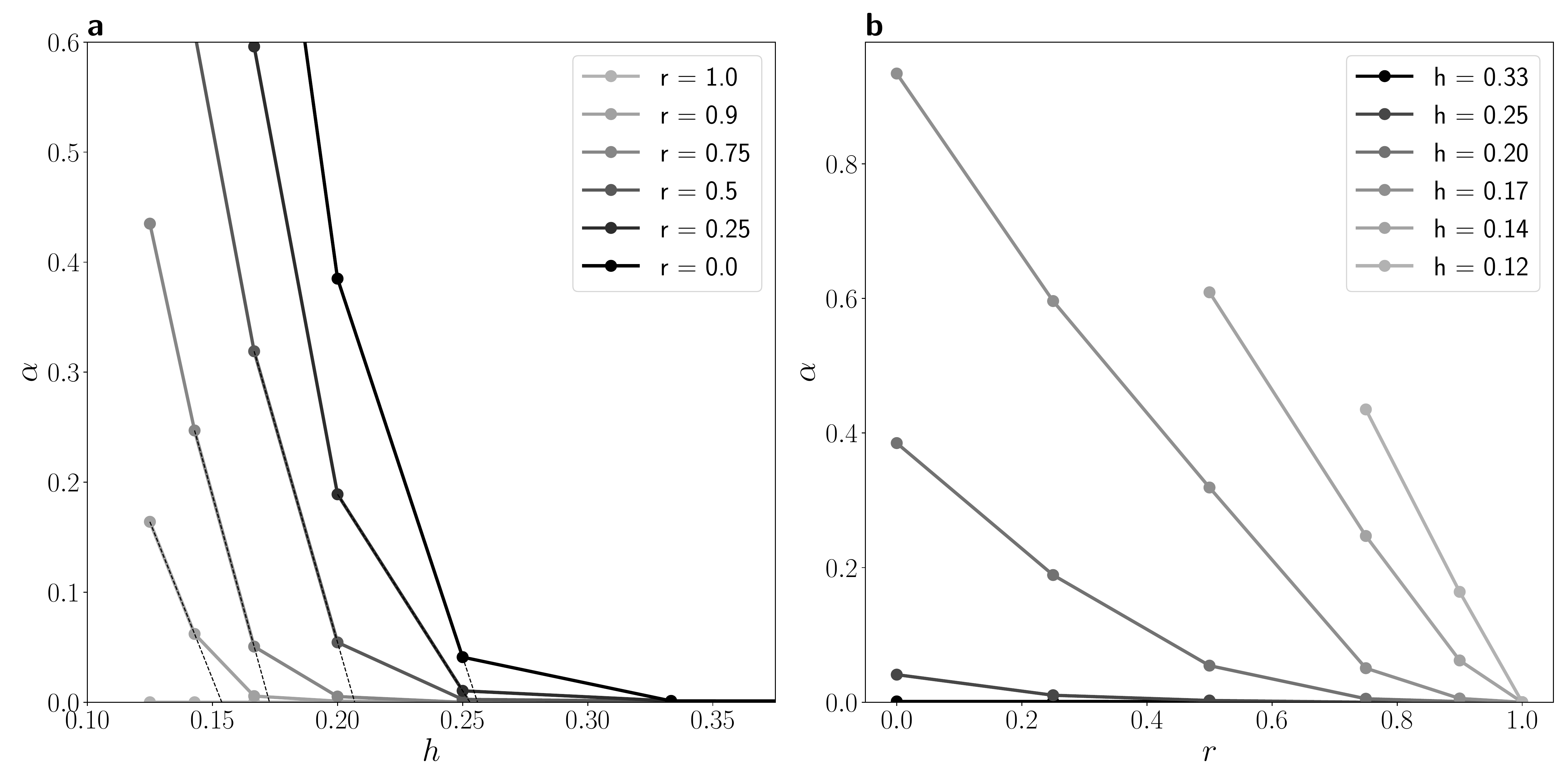}
\end{center}
\caption{Strength of the inverse cascade for different values of $r$ and $h$, shown as a function of $h$ (panel \textbf{a}) and $r$ (panel \textbf{b}). All these values correspond to the set of simulations L2Re1. Thin dashed lines show how the extrapolation is made for a hypothesis of the critical value. \label{fig:alpha}}
\end{figure*}

To determine if the system is in a state where an inverse cascade is present, we measure the relative amplitude $\alpha$ as a function of the height of the layer $h$ and the forcing dimensionality parameter $r$. 
Figure \ref{fig:alpha} shows $\alpha$  as a function of $r$ for different values of  $h$ 
in the left panel, while the right panel shows $\alpha$ as a function of $r$ for different values of $h$. The results are from the series of runs L2Re1. For all values of $r$, the strength of the inverse cascade is decreasing with $h$ (see left panel). Similarly, for all values of $h$, $\alpha$ is decreasing with $r$ (see right panel). This is consistent with our expectations and with previous studies, both for the layer thickness in \cite{benavides2017,celani2010} and for the dimensionality of the forcing suggested by the amplitude of the negative eddy viscosity in \cite{alexakis2018b}. It is worth noting that the case $r=1$ (i.e. fully 3D forcing) does not produce any inverse cascade. This demonstrates the crucial influence of the characteristics of the forcing on the behavior of the flow in the inertial range. We cannot be certain, of course, that this result persists as $\Lambda$ and $Re_\mu,Re_\nu$ are increased. 

For any value of $r$ different from $1$, the transition from a split cascading case ($\alpha>0)$ to  strictly forward case $\alpha \simeq 0$ is observed around a value of of $h$ that we denote as $h_c(r)$.  One can clearly observe that the value of $h$ below which an inverse cascade appears decreases with $r$ (figure \ref{fig:alpha}a). Conversely, it also seems that for each value of $h$, there is a value $r$ separating the existence from the absence of an inverse cascade (figure \ref{fig:alpha}b). The transition value of $h$ at $r = 0$ is very close to the one found by \cite{benavides2017} for which the same forcing was used, around $h = 1/4$, despite the small differences between the equations solved here and in those works. 


Whether this transition is sharp, i.e. whether there is a critical value of $h=h_c(r)$ above which $\alpha$ is exactly zero, cannot be concluded from the present data. However, based on previous studies we do expect that the transition will become sharper as $\Lambda$ and $Re_\mu,Re_\nu$ are increased, converging to a critical transition. We examine this particular limiting procedure for $r=0.5$ in the next section. Since this is an computationally expensive procedure to follow for each value of $r$, we will limit ourselves to estimating the value of $h_c(r)$ by extrapolation: For a given value of $r$, we linearly extrapolate the last non-zero values of $\alpha(h,r)$ (see left panel of fig. \ref{fig:alpha} to obtain the value of $h_c(r)$ for which $\alpha(h_c,r)$ is zero).

\begin{figure}[h]
\begin{center}
\includegraphics[width=9cm]{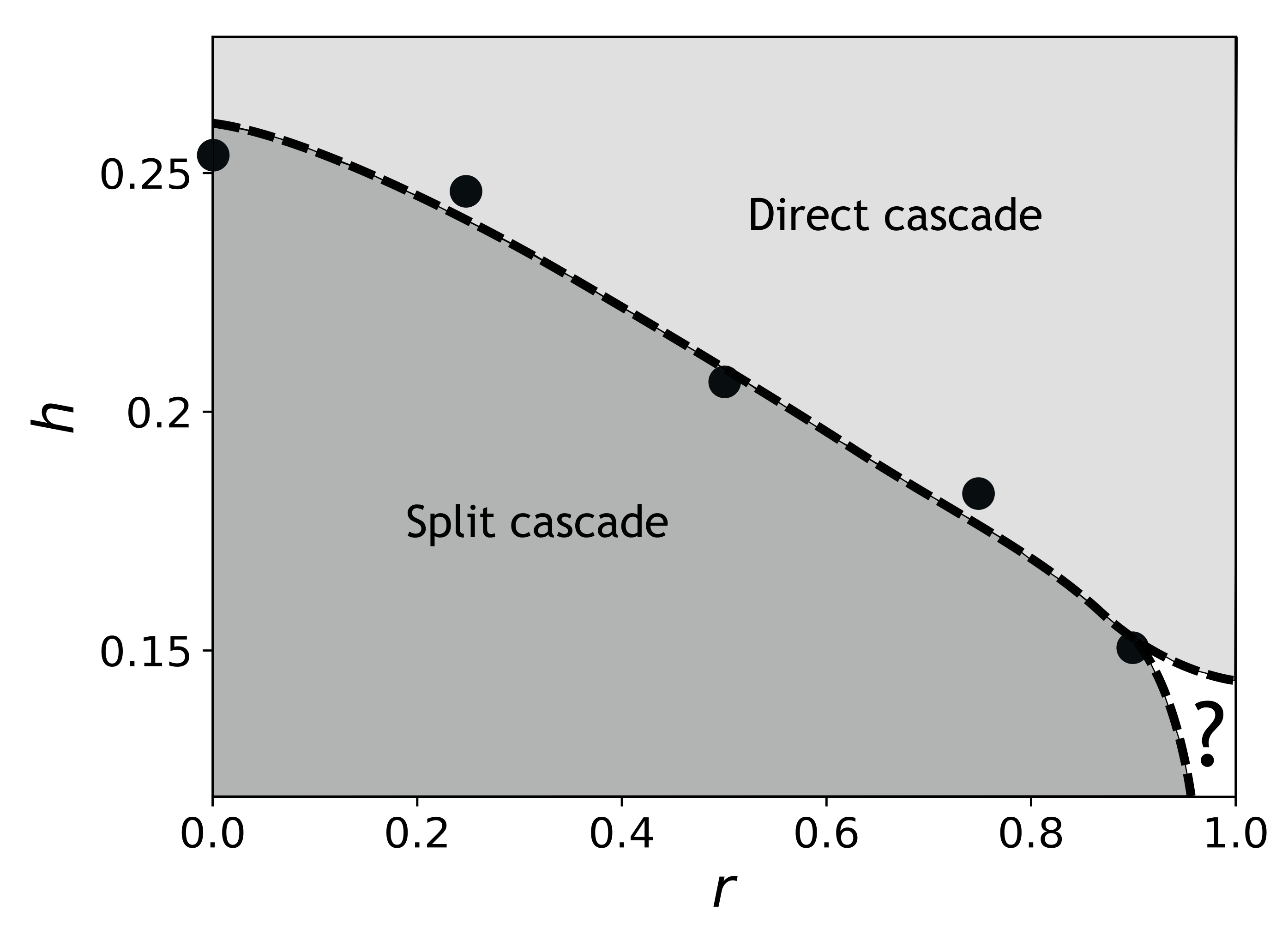}
\end{center}
\caption{Schematic of the regions in $r-h$ space where an inverse cascade occurs or not. The dots show the critical values estimated from this study, and the line is a hypothesis for the shape of the diagram limit. \label{fig:phase}}
\end{figure}
This enables us to draw a phase diagram showing the critical height $h_c$ as a function of $r$ (or conversely). The curve separates the two different observed phases: a split cascade for $h<h_c(r)$ and pure direct cascade for $h>h_c(r)$. For $r$ very close to unity, we cannot be certain for the behavior of $h_c(r)$. If the $h_c(r)$ does indeed tend to 0 for $r\le1$ would imply that the $r=1$ case cannot generate a split cascade, no matter how thin the layer. On the other hand, if $h_c(1)$ is finite, this would imply that even the case $r=1$ can generate an inverse cascade if the layer is thin enough.
We leave this open issue for future research.

\begin{figure}
\begin{center}
\includegraphics[width=17cm]{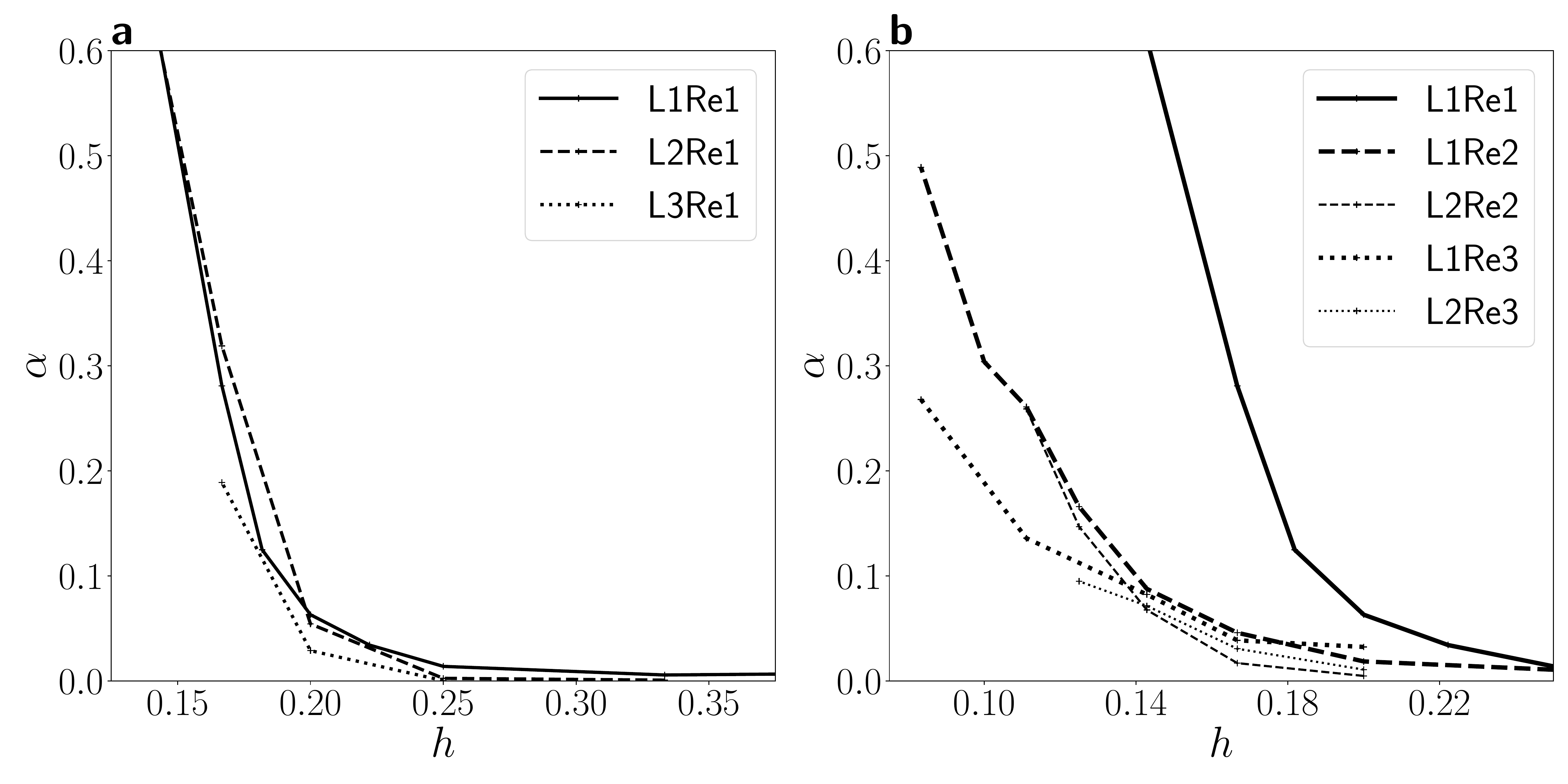}
\end{center}
\caption{Strength of the inverse cascade for different values of $h$, shown as a function of $h$. Different line styles correspond to different sets of simulations, as indicated in the legends. All the simulations were done with the parameter $r=0.5$. \textbf{a} : Behavior in the large-box limit. \textbf{b} : Behavior in the low-viscosity limit. \label{fig:conv}}
\end{figure}

\subsection{Convergence} 
As stressed in section \ref{CPO} (see also \cite{alexakis2018}), it is important to examine the limit $\lambda\to \infty$ and $Re_\mu,Re_\nu\to \infty$. However, realising such a limiting procedure requires significant computational time, and therefore this limit was pursued only for a few values of $r$.
In the left panel of figure \ref{fig:conv}, we show $\alpha(h)$ for $r=0.5$ for $\Lambda=8$, $\Lambda=16$ and $\Lambda=32$. Indeed, the transition from a direct to a bidirectional cascade becomes sharper as the box size is increased. This suggests that in the large-box limit the curve will converge to a sharp transition with a critical value $h_c(r)$ that separates the pure direct cascade regime $\alpha=0$ from the bidirectional regime. The same behavior is found for all the values of $r$ examined (but only $r=0.5$ is shown in figure \ref{fig:conv}a). 

In the right panel of figure \ref{fig:conv}, we show results from the simulations for different values of $Re_\nu$ and $\Lambda,Re_\mu$.
An increase of $Re_\nu$ appears to decrease the value of $h_c$ where the transition takes place:
as the Reynolds number is increased, the curves move to the left. The critical height $h_c(r)$ does decrease between the simulations L1Re1 and L1Re2. A similar tendency was observed in the results of \cite{vankan2019}, who also found that the critical value of $h$ was decreasing with the Reynolds number in the simulation. However, as the box size is increased, 
this dependency seems to diminish. For example, the critical value of $h$ does not decrease between the simulations L2Re2 and L2Re3. Therefore, there seems to be convergence of the critical value $h_c(r)$ with increasing Reynolds number and increasing box size and will become independent of 
$\Lambda,Re_\mu,Re_\nu$. However, future studies at higher resolutions need to verify the observed tendency that is noted here. In practical terms, these results imply that the results in figure \ref{fig:phase} are likely to overestimate the value of $h_c(r)$ and their true value, valid in the 
large $Re_\nu$ limit, is probably smaller.

\section{Conclusions}
\label{sec:conclusions}

This study investigated the role of the forcing in the turbulent energy cascade of a flow. 
It was demonstrated that the forcing, and in particular its dimensionality, can change the properties of turbulence and even the direction of the energy cascade. 
In particular, it was shown that, for the particular choice of forcing parameterization used,
the critical height $h_c$, where the transition from forward to a split cascade takes place,
depends of the forcing dimensionality parameter $r$. The more `3-D' the forcing is, the thinner
the layer is required to be for an inverse cascade to appear. Within some approximation, we were able to track the dependence of $h_c$ on $r$ and thus we were able to construct a phase space diagram that gives the locations in the $(h,r)$ plane where a split cascade is met. A decrease of $h_c$ with increasing Reynolds number was observed, implying that our estimation of $h_c$ is probably slightly 
higher than the true value.

There are several open questions that are left for future work. First of all,
in this work we studied a very limited parameterization of the forcing that consisted of a linear combination of 2-D eddies and 3-D convection cells. In natural flows there is a wide range of mechanisms that can inject energy to a flow typically consisting on some instability such as convection. It is thus hard to extrapolate the present results to such forcing mechanisms.
However, we do believe that the tendency that three-dimensionality of the forcing suppresses 
the inverse cascade should be a general result. Different mechanisms however need to be explored independently. 

A particular result obtained in this work was that no inverse cascade was observed  for $r=1$.  This implies that when our forcing is purely 3D and takes the form of convection cells, then no inverse cascade is present. This was true for all box sizes and $Re_\nu,Re_\mu$ examined. This result needs to be verified at larger resolutions, but if true, it would suggest that in convection in the absence of rotation, no inverse cascade will be observed (see for example \cite{emran2015large}), as opposed to the case of rotating convection \cite{favier2014inverse,guervilly2014large,rubio2014upscale}. 

Clearly, this study only focuses on 
the role of the forcing in a particular configuration and for a particular problem. Other studies, looking at the influence of other parameters of the forcing (such as the injection scale, helicity etc.) might be useful in order to assess to what extent the forcing determines the properties of turbulence, on top of the already well-known influence of  other external parameters.

\acknowledgments 
This work was granted access to the HPC resources of MesoPSL financed by the Region 
Ile de France and the project Equip@Meso (reference ANR-10-EQPX-29-01) of the programme Investissements d'Avenir supervised by the Agence Nationale pour la Recherche and the HPC resources of GENCI-TGCC \& GENCI-CINES (Project No. A0070506421) where the present numerical simulations have been performed. This work has also been supported by the Agence nationale de la recherche (ANR DYSTURB project No. ANR-17-CE30-0004).

\bibliography{report,transitions}
\end{document}